**Test of a theory of the Mott quantum-measurement problem**

Jonathan F. Schonfeld
Center for Astrophysics | Harvard and Smithsonian
60 Garden St., Cambridge, Massachusetts 02138 USA
jschonfeld@cfa.harvard.edu
ORCID ID# 0000-0002-8909-2401

**Abstract** The Mott problem asks: Is there a microphysical mechanism – based only on Schroedinger's equation – that explains why an alpha particle emitted in a spherically symmetric nuclear decay produces a non-spherically-symmetric single track in a cloud chamber? This is a variant of the more general quantum measurement problem. Earlier, I proposed such a mechanism, drawing on quantum-mechanical Coulomb scattering and the thermal behavior of supersaturated vapors. I found that the probability that a track originates at distance $R$ from the decay source is proportional to $1/R^2$, with a proportionality constant that I expressed in terms of more fundamental parameters but was unable to estimate at the time. I tested the $1/R^2$ law opportunistically using cloud chamber video from the Internet. Here, I draw on chemical physics to independently estimate the proportionality constant. The estimate is within a factor 1-2 of a value extracted directly from the data.

**Funding**

The author declares that no funds, grants, or other support were received during the preparation of this manuscript.

**Competing interests**

The author has no relevant financial or non-financial interests to disclose.

## 1. Introduction

Canonical quantum mechanics consists of a smooth evolution law (Schroedinger's equation) that acts on a wavefunction, plus rules that prescribe what happens when a measurement interrupts smooth wavefunction evolution. One might expect that measurement rules would be implicit in the evolution law itself (supplemented by a sufficiently careful and self-consistent model of a measurement apparatus), but that has remained unproved for almost a hundred years [1]. So one has grown accustomed to viewing the measurement rules as axioms in their own right, logically partitioned from anything else in the theory. The rules are familiar:

- A measured quantity corresponds to a Hermitian operator.
- A measurement of that quantity projects the state being measured onto a single eigenvector of this operator.
- The specific single eigenvector is a random outcome.
- The probability of this random outcome is governed by the Born rule: it's the square absolute-value of the Hilbert-space overlap between the incoming wavefunction and the eigenvector in question.

The measurement rules for cloud chamber detection of an alpha particle from an s-wave nuclear decay are similar but not identical:

- It's not obvious what the relevant Hermitian measurement operator is.
- Detection collimates the original spherically symmetric alpha wavefunction into a *single* thin straight track.
- The specific direction and starting point of the single track is a random outcome.
- The tradition does not dictate a preferred radial probability distribution for the starting point (the distribution of its solid angle is presumably spherically symmetric).

In the general case, the challenge of showing that measurement rules are implicit in smooth evolution is referred to as the quantum measurement problem. In the cloud chamber case, it's referred to as the Mott problem. This honors the 1929 paper [2] in which N. Mott used the Schroedinger equation to explain why alpha particles in cloud chambers should have track-like signatures. The 1929 paper did not explain why or how one observes only one track at a time, or what its probability distribution should be.

One studies cloud chambers because their underlying physics is relatively simple. And it seems intuitive that if one could show how measurement rules are implicit in smooth evolution in that restricted case, i.e., resolve the Mott problem in the affirmative, that would be good progress toward the more general measurement problem as well.

The present paper extends work on the Mott problem begun in Reference [3] and continued in Reference [4]. In Reference [3], I developed a theory of the Mott problem, building on basic facts about the long-time behavior of the alpha-particle wavefunction, and drawing on details of quantum-mechanical Coulomb scattering and the thermal behavior of supersaturated vapors. I concluded that, in a large enough sample, the probability that a cloud chamber track originates at distance $R$ from the decay source is proportional to $1/R^2$ (a kind of Born rule), with a proportionality constant which I expressed in terms of more fundamental parameters involving diverse physical processes, but which I was unable to estimate at the time from first principles. In Reference [4], I tested the $1/R^2$ law opportunistically using pedagogical cloud chamber video posted on the Internet. I found plausible consistency out to a limiting value of $R$, and speculated about what's going on beyond that apparent cutoff. Since then, I became aware of work in the chemical physics literature [5] that enables a rough but independent first-principles estimate of the proportionality constant. That is the subject of the present paper. The estimate is quite rough, but within a factor 1-2 of a value that I extract directly from the video data. Given the crudeness of the experimental data, the roughness of the numerical estimates, and the extreme spread of concentrations involved (air molecules [$2.5\times10^{16}$ mm$^{-3}$] vs. subcritical vapor clusters of specific numbers of constituents [~$8\times10^6$ mm$^{-3}$, see Equation (3.20)]), this is significant agreement at this stage in the theory's development.

We review the salient aspects of the earlier work in Section 2, below. We estimate the coefficient of $1/R^2$ in Section 3. Section 4 contains concluding remarks.

## 2. Review of prior work

[This and the next three paragraphs paraphrase content from Reference [3].] A diffusion cloud chamber is an enclosure containing air commonly supersaturated with ethyl alcohol. Degree of supersaturation around 6 is typical [6]. The chamber is cooled from below, so supersaturation matters only in a layer at the bottom. When a charged particle passes through the sensitive layer, it ionizes air molecules, and the ions nucleate visible vapor droplets (drive vapor molecule clusters supercritical). This picture is easy to understand when the charged particle wavefunction is strongly collimated (so the particle can be treated as a point at one location moving in one direction). But the actual wavefunction of an alpha particle near the source of an s-wave radioactive decay is not collimated in any meaningful sense. As a decay product, the alpha occupies a Gamow state [7], an eigenfunction of the Hamiltonian with complex eigenvalue (this was called "persistent state" in Reference [3]). Despite the complex eigenvalue, a Gamow state is normalizable because it depends on time via the usual multiplier exp(-i$Et$) supplemented by a cutoff at a wavefront that expands according to the classical speed of the outgoing alpha particle. The alpha Gamow-state wavefunction near the decay source at three-dimensional position **x** and time $t$ is given by

$$\psi(\boldsymbol{x},t) \sim -i\sqrt{\frac{\gamma}{4\pi v}}\frac{1}{R}exp\left\{\left(\frac{R}{v}-t\right)\left(\frac{\gamma}{2}+i\frac{Mv^2}{\hbar}\right)\right\}, \tag{2.1}$$

where **x** is defined relative to the location of the initial decaying nucleus, $R\equiv\|\mathbf{x}\|$ is distance from the initial heavy particle, $M$ is emitted particle mass, $v$ is emitted particle speed, $\gamma$ is the conventional decay e-folding rate [i.e. ln(2) divided by half-life], and we ignore an irrelevant overall phase factor exp(i[total energy]$t/\hbar$). (This also corrects a sign error in [3].) Strictly speaking, Equation (2.1) holds outside the decay source's interaction radius, which we ignore because we're interested in length scales characteristic of the cloud chamber, while the interaction radius is on a nuclear scale. Equation (2.1) ignores effects due to the cloud-chamber enclosure walls.

The wavefunction of Equation (2.1) interacts with an already existing vapor cluster (generated randomly due to thermal fluctuations) that is just barely sub-critical. A barely sub-critical cluster turns out to have a very large amplitude of interaction with the wavefunction of Equation (2.1), so that even a very weak wavefunction can provoke the subcritical droplet to grow quickly in a supercritical fashion and become visible, and provoke the alpha wavefunction to collimate. This is because, in the presence of a cluster that's already formed, single-molecule ionization can proceed with very small energy loss, since ion-induced potential energy due to droplet polarization can nearly balance electron excitation energy (a form of Penning ionization [8]). This near-degeneracy drives the cross section of this quantum Coulomb interaction to singularity, and leads to a collimated nuclear-decay-product free wavefunction in the outgoing state.

Within this picture, the probability per unit time and unit volume to find an emitted-particle track originating at three-dimensional position **x** and time $t$ is approximately

$$\mathrm{P}(\mathbf{x},\mathrm{t}) \sim \rho \tau A v |\psi(\boldsymbol{x},t)|^2, \tag{2.2}$$

where (see next section) the density $\rho$ and time $\tau$ are constants characteristic of the cloud chamber medium, $A$ is a constant characteristic of the ionization process, and time $t$ is measured from the establishment of the source ($\rho$, $\tau$ and $A$ will all be defined in detail in the next section). Equation (2.2) is a Born rule, in the sense that it asserts a proportionality between a measurement probability and the squared absolute value of a wavefunction in a particular coordinate system. Substituting Equation (2.1) for $\psi$, and assuming small $\gamma$ (slow decays), this reduces to

$$\mathrm{P}(\mathbf{x},\mathrm{t}) = \left(\frac{\gamma \rho \tau A}{4\pi}\right) e^{-\gamma t} \frac{1}{R^2}. \tag{2.3}$$

Obviously, if $N$ is the total number of nuclei confined to a very small volume, then the total number of track originations detected per unit volume and unit time is $N$P(**x**,$t$).

[Equation (2.1) is naïve in at least one important respect. As discussed in Reference [9], the alpha wavefunction is actually "shredded" by virtual inelastic encounters with the molecules in the ambient cloud chamber medium. We should expect that, as a result, the wavefunction at increasing $R$ represents an alpha particle of diminishing speed. That should not affect the simple $1/R^2$ proportionality in Equation (2.3), because the factor $A$, which describes the cross-section for alpha-induced ionization near threshold, turns out to be independent of alpha speed $v$, as we show in the Section 3.]

Reference [4] reported a first attempt to test Equation (2.3) with experimental data. Since, at the time, I had no basis for estimating the factor $\rho\tau$ from first principles, I confined myself to testing the inverse-square dependence on $R$. I could (and still can) find no record of experiments specifically aimed at measuring the statistics of cloud-chamber track origination from alpha decay. Lacking that, I sought and located a suitable video of opportunity [10] on the Internet, and tabulated instances of track origination by stepping through the video frame by frame and extracting coordinates by hand. The cloud chamber in the video was enclosed in a standard Petri dish, and I observed 228 track starts originating out to roughly 22.5mm from the radioactive source in the center. I observed none beyond 22.5mm. Allowing for the crudity of the underlying data, I found plausible agreement with a $1/R^2$ law within 22.5mm. I explored several possible explanations for the apparent cutoff at 22.5mm, ruling out some but not others. In the present paper I am careful to confine the analysis below to the space $R$<22.5mm.

After publication of Reference [4], I became aware of applicable *a priori* estimates of what amounts to $\rho\tau$ in Reference [5]. In the next section I attempt to test Equation (2.3) by comparing these estimates with a value that I extract from the data in Reference [4].

## 3. Test of $1/R^2$ proportionality constant

The narrative here is divided into subsections describing individual steps that build to the test in question.

*Measured value of $(N\gamma\rho\tau A)e^{-\gamma\tau}$*

As discussed in Reference [4], the diffusion cloud chamber in video [10] has a sensitive layer of thickness ~4mm, and the radioactive point source is located 3.5mm above the layer's floor. As a result, according to Reference [4], the total number of counts (=228) observed within 22.5mm during the video's 90 sec of cloud chamber observation should satisfy

$$228 = (90\text{s})\left(\frac{N\gamma\rho\tau A}{4\pi}\right)e^{-\gamma t}[C(R,b) + C(R,a-b)], \tag{3.1}$$

where

$$C(R,b) \equiv \pi b \ln\left[1 + \left(\frac{R}{b}\right)^2\right] + 2\pi R \cdot Arctan\left(\frac{b}{R}\right) \tag{3.2}$$

and $R$=22.5mm, $b$=3.5mm and $a$=4mm. With this definition and these parameters, we have

$$[C(R,b) + C(R,a-b)] = 78mm \tag{3.3}$$

and therefore

$$(N\gamma\rho\tau A)e^{-\gamma t} = 0.4mm^{-1}s^{-1}. \tag{3.4}$$

*Estimate of N*

The radioactive emitter in video [10] is a 0.01μCi $^{210}$Pb "needle source" (sample plated to the tip of a sewing needle) from Spectrum Techniques of Oak Ridge TN [11]. Per its decay chain, the $^{210}$Pb is admixed with $^{210}$Bi and $^{210}$Po. $^{210}$Pb and $^{210}$Bi are beta emitters (half-lives 22.3 years and 5.0 days, respectively), and $^{210}$Po is an s-wave alpha emitter (138 days) [12]. Only the alpha tracks are visible in the video [10], so *N* is the size of the Polonium population. Given the hierarchy of decay times, we can assume that the polonium population fraction is roughly the ratio of the $^{210}$Po to $^{210}$Pb lifetimes, i.e. 1.7%; the bismuth fraction is even smaller. We can also assume that 1/3 of the 0.01 μCi represents the alpha flux, and the other two thirds represent the (unseen) beta fluxes of $^{210}$Pb and $^{210}$Bi.

Since each $^{210}$Pb nucleus is ultimately responsible for three decay products (two betas and one alpha) the total $^{210}$Pb population $N_T$ must satisfy

$$\frac{3N_T}{(22.3\ yr)} \sim 0.01\mu Ci = 370\frac{counts}{s} \tag{3.5}$$

and therefore

$$N \sim 0.017 N_T \sim 1.5 \times 10^9. \tag{3.6}$$

*Estimate of A*

The factor *A* in Equations (2.2) and (2.3) characterizes the cross section for an alpha particle to ionize a vapor atom near a Penning threshold. Following Reference [3] (but modifying a few pieces of notation), let *r* be the radius of a subcritical vapor cluster that contains the atom to be ionized. The atom, when ionized, induces a negative energy of polarization in the cluster, and this tends to compensate for the binding energy of the electron that's ejected when the ion forms (Penning ionization process). Let $r_c$ be the critical radius at which the energy of polarization exactly balances the electron binding energy (Penning threshold). Then, for *r* slightly less than $r_c$, the total cross-section for alpha-induced ionization takes the form $A/(r_c - r)$.

To understand how to estimate *A*, at least at the order-of-magnitude level, start by considering the fundamental theoretical expression for differential ionization cross section in the distorted-wave Born approximation from Reference [3]:

$$d\sigma = 8\pi S Q^2 v^{-2} |F(\boldsymbol{q}, E_e, d\Omega_e)|^2 q^{-3} dq dE_e d\Omega_e, \tag{3.7}$$

where *F* is a sum of atomic-state overlap integrals; $E_e$ and $\Omega_e$ are outgoing electron kinetic energy and solid angle, respectively; $\hbar\mathbf{q}$ is the difference between outgoing and incoming alpha momentum; and *Q* and *v* are alpha charge and speed, respectively, assuming small energy loss.

[*S* is the multiplicity of the atomic shell that furnishes the ejected electron (this corrects an omission in Reference [3]). For Carbon, with two electrons in the 2p shell, *S* would naively be 2;

for Oxygen, with four 2p electrons, $S$ would naively be 4. However, since my account of the Mott problem relies on close matching between ionization potential and the energy of polarization induced in a vapor cluster, it seems wrong to take intra-shell degeneracy for granted. Thus, in what follows we ignore $S$ by implicitly setting it to unity.]

For small $E_e$, we ignore the dependence of $F$ on kinematic variables, so the total ionization cross section becomes

$$\sigma \sim 8\pi v^{-2}|QF|^2 \int q^{-3} dq dE_e d\Omega_e = 2|4\pi QF|^2 v^{-2} \int q^{-3} dq dE_e. \tag{3.8}$$

The d$E_e$ integral is the electron energy spread,

$$\int dE_e = \frac{1}{2M}(p+q)^2 - \frac{1}{2M}p^2 \sim \frac{pq}{M} = vq, \tag{3.9}$$

where $M$ is alpha mass and $p$ is incoming alpha momentum. The minimum possible value of $q$, $q_{\min}$, satisfies

$$E_0 = \frac{1}{2M}(p+q_{min})^2 - \frac{1}{2M}p^2 \sim \frac{pq_{min}}{M} = vq_{min}, \tag{3.10}$$

where $E_0$ is electron ionization potential. Thus, combining the last three equations, we have, in the limit of very small $E_0$,

$$\sigma \sim 2|4\pi QF|^2 v^{-2} \int q^{-3} dq dE_e \sim 2|4\pi QF|^2 v^{-1}(1/q_{min}) \sim \frac{2|4\pi QF|^2}{E_0}. \tag{3.11}$$

(As promised, the $v$-dependence has dropped out.) The numerator in Equation (3.11) has dimensions [area]×[energy]. For the area, our rough estimate is $\pi r_a^2$, where $r_a$ is atomic radius. Since the cloud chamber atoms most likely to ionize (C, N, O) all have outer electrons in the $n$=2 shell, we estimate $r_a \sim 4\times$Bohr radius$\equiv 4r_B$. For the energy, we estimate 1 Rydberg. One might have expected [1 Rydberg]/$2^2$, but dividing by 4 is compensated for by the implicit factor of 2 in the alpha charge $Q$.

To complete our estimate of $A$, we need to relate the denominator $E_0$ to $r_c$ - $r$. Since $E_0$ is a small difference between electron binding energy and the polarization energy induced in a cluster; and since [3] the polarization energy depends on cluster radius $r$ as [constant-$e^2/2r$] for large cluster dielectric constant, we can write ($e$ is electron charge)

$$E_0 \sim (r_c - r)\left[\frac{d}{dr}\left(-\frac{e^2}{2r}\right)\right]_{r=r_c} = (r_c - r)\frac{e^2}{2r_B}\left(\frac{r_B}{r_c^2}\right) = (r_c - r) \times (1\ Rydberg) \times \left(\frac{r_B}{r_c^2}\right). \tag{3.12}$$

To estimate $r_c$, we first invoke Reference [13] to estimate the number of molecules in a critical cluster as $u$~25. Let us further estimate the volume per molecule as $d^3$, where [14] $d$~(liquid ethanol density)$^{-1/3}$~10$r_B$. Then

$$r_c = d\left(\frac{3u}{4\pi}\right)^{1/3} \sim 18r_B. \tag{3.13}$$

Putting together Equations (3.11), (3.12) and (3.13), we have, finally,

$$\sigma \sim \frac{(1\ Rydberg) \times 16\pi {r_B}^2}{(r_c - r) \times (1\ Rydberg) \times \left(\frac{1}{18^2 r_B}\right)}. \tag{3.14}$$

It follows (using $r_B$~0.5 Angstrom) that

$$A \sim 5{,}200\pi {r_B}^3 \sim 2 \times 10^{-18} mm^3. \tag{3.15}$$

*Implied value of $\rho\tau$*

It is now elementary to combine Equations (3.4), (3.6) and (3.15) to arrive at

$$\rho\tau e^{-\gamma t} \sim 2 \times 10^{15}\ mm^{-4}, \tag{3.16}$$

where we have used $\gamma$=(polonium half-life)$^{-1}$×ln2 =6×10$^{-8}$s$^{-1}$. Since the polonium population is continually replenished by the slow $^{210}$Pb decay, it's reasonable to set $\gamma t$~1, from which it follows that

$$\rho\tau \sim 5 \times 10^{14}\ mm^{-4}. \tag{3.17}$$

*A priori estimate of $\rho\tau$*

Reference [3] defines $\rho$ as the number of subcritical vapor clusters formed per unit volume and unit time, and unit interval of cluster radius $r$. But this is not quite right, because I used it as if it were the number of relevant ionizable atoms contained in subcritical clusters, and there are 25 vapor molecules per cluster and 1 or 2 (O or C) ionizing atoms per molecule. So in what follows we replace Equation (3.17) with

$$\rho\tau \sim \frac{5 \times 10^{14}}{K} mm^{-4}, \tag{3.18}$$

where $K$ is 25 or 50. (In an earlier preprint of this paper, I had failed to recognize the need for the factor $K$. This still needs to be viewed with caution, because the various atoms counted by $K$ are in different locations within the cluster and therefore have different energetic environments.) Reference [3] also defines $\tau$ as the time for such clusters to evaporate. Thus we can interpret the product $\rho\tau \equiv \rho_{ss}$ as the number of clusters in metastable steady state per unit volume and per unit interval of cluster radius $r$. This is important because Reference [5], to which we want to benchmark the empirical estimate in Equation (3.18), concerns itself with steady-state densities. However, the densities in Reference [5] are indexed by $u$, the number of molecules in a cluster, rather than by the cluster's radius $r$, but we can translate easily. Since $(4/3)\pi r^3=ud^3$, the difference in $r$ between two clusters of populations $u$ and $u$+1 satisfies

$$\delta r = \frac{d}{(36\pi u^2)^{\frac{1}{3}}}, \tag{3.19}$$

and therefore the density $\rho_u$ of clusters of population $u$ is related to $\rho_{ss}$ by

$$\rho_u = \left[\frac{d}{(36\pi u^2)^{\frac{1}{3}}}\right]\rho_{ss}. \tag{3.20}$$

Estimating, as before, $d$~$10r_B$ and $u$~25, this means that Equation (3.18) is equivalent to

$$\rho_u \sim \frac{1.2\times 10^7}{K} mm^{-3} \sim \frac{2\times 10^{-11}}{K} mol/liter. \tag{3.21}$$

This is what we need to compare with the theoretical results in Reference [5], and we have adjusted the units accordingly.

We now turn to selecting the appropriate results from Reference [5] for comparison. Reference [5] reports molecular-kinetics calculations of the densities of clusters of various sizes, for homogeneous gases $H_2O$, cesium and n-pentanol at distinct temperatures and supersaturation 14.84, 7.57 and 3.442, respectively,. Strictly speaking, none of these is the cloud chamber scenario, which concerns a mixture of two gases, one condensing (alcohol) and one not (air). Moreover, molecular-level calculations of cluster formation have made considerable progress since Reference [5] was written, but more recent work also focuses on homogeneous systems [15, 16] or assumes extreme supersaturation [17]. Nonetheless, this will be our basis for extrapolating from numerical results in Reference [5] in order to benchmark Equation (3.21). Our target cloud-chamber supersaturation, 6, is between the cesium and n-pentanol cases.

Reference [5] Table II reports results for supersaturation 7.57. The table shows $\rho_u$ for $u$ ranging from 51 to 105. [Reference [5] uses the notation $A_{u,m}$ for $\rho_u$, where the integer index $m$ is a kind of computational cutoff.] $\rho_u$ gains a factor of $2\times10^8$ when $u$ decreases by 23, from 74 to 51, and there seems to be a rough exponential trend otherwise. So if $\rho_u$ is $4\times10^{-21}$ mol/liter for $u$=51, it extrapolates to ~$10^{-11}$ mol/liter for the target $u$=25. This may be an overestimate, since the same extrapolation produces $\rho_{u=1}$~$10^{-1}$ mol/liter, while the caption to Table II of Reference [5] indicates $\rho_{u=1}$~$2.5\times10^{-4}$ mol/liter.

Reference [5] Table III reports results for supersaturation 3.442. There, $\rho_u$ gains a factor of $1.6\times10^9$ when $u$ decreases by 21, from 70 to 49. So if $\rho_u$ is $7\times10^{-26}$ mol/liter for $u$=49, it extrapolates to $\sim2\times10^{-15}$ mol/liter for the target $u$=25. This may be an underestimate, since the same extrapolation produces $\rho_{u=1}\sim2.5\times10^{-5}$ mol/liter, while the caption to Table III of Reference [5] indicates $\rho_{u=1}\sim1.9\times10^{-3}$ mol/liter.

To hypothesize a value for supersaturation 6, I exponentially average these two estimates and get $\rho_u\sim4\times10^{-13}$ mol/liter (i.e. I weight the two estimates' logarithms according to the fraction that 6 represents of the distance from 3.442 to 7.57). This interpolation appears to match the empirical estimate in Equation (3.21) for $K$=50, and to be within a factor of 2 for $K$=25. This is the principal result in this paper.

## 4. Remarks

The principal result in this paper, combined with the results in Reference [4], represents support for the theory of the Mott problem embodied in Equation (2.3). Reference [4] addressed the functional form of Equation (2.3), and the present paper addresses its overall normalization. Of course this demonstration is no stronger than the underlying data, which come from an uncontrolled experiment not designed for this purpose. This demonstration is also limited by the quality of our numerical estimates, which are exceedingly rough. And, finally, it is also limited by the quality and relevance of the computations in Reference [5]: a simulated gas of pure cesium with a computational cutoff is not the same as a real sample of air supersaturated with ethanol. Nevertheless, in view of the extreme spread of concentrations involved (air molecules vs. subcritical vapor clusters of specific sizes), I think the new result is worth paying attention to. Further work will have to include more careful experimentation (real and computational) on the statistics of both charge-particle tracks and subcritical vapor clusters in cloud chambers.

## References

1. Allahverdyan, A. E., Balian, R., Nieuwenhuizen, T. M.: Understanding quantum measurement from the solution of dynamical models. Physics Reports. **525**, 1-166 (2013). https://doi.org/10.1016/j.physrep.2012.11.001
2. Mott, N.: The wave mechanics of $\alpha$-ray tracks. Proc. R. Soc. Lond. A (1929). https://doi.org/10.1098/rspa.1929.0205
3. Schonfeld, J. F.: The first droplet in a cloud chamber track. Found. Phys. **51**, 47 (2021). https://doi.org/10.1007/s10701-021-00452-x
4. Schonfeld, J. F.: Measured distribution of cloud chamber tracks from radioactive decay: a new empirical approach to investigating the quantum measurement problem. Open Physics 20, 40 (2022). https://doi.org/10.1515/phys-2022-0009
5. Bauer, S. H., Zhang, X.-Y., Wilcox, C. F.: Metastable size distributions of molecular clusters in supersaturated vapors. J. Chem. Phys. **114**, 9408 (2001). https://doi.org/10.1063/1.1370076
6. Mori, C.: Visibility of the growth direction of an alpha-particle track in a diffusion cloud chamber. Journal of Nuclear Science and Technology (2014). https://doi.org/10.1080/00223131.2014.854710

7. Schonfeld, J. F.: Generalization of Gamow states to multi-particle decay. arXiv:2204.05854 [quant-ph], (2022). http://www.arxiv.org/pdf/2204.05854
8. Efimov, D. K., Miculis, K., Bezuglov, N. N., Ekers, A.: Strong enhancement of Penning ionization for asymmetric atom pairs in cold Rydberg gases: the Tom and Jerry effect. J. Phys. B: At. Mol. Opt. Phys. **49**, 125302 (2016). https://doi.org/ 10.1088/0953-4075/49/12/125302
9. Gaspard, D., Sparenberg, J.-M.: Solvable model of a quantum particle in a detector. International Journal of Quantum Information **17**, 1941004 (2019). https://doi.org/ 10.1142/S0219749919410041
10. https://www.youtube.com/watch?v=pewTySxfTQk Jefferson Lab, 10/25/2010.
11. https://www.spectrumtechniques.com/products/sources/needle-sources/
12. http://nucleardata.nuclear.lu.se/toi/nuclide.asp?iZA=820210, http://nucleardata.nuclear.lu.se/toi/nuclide.asp?iZA=830210, http://nucleardata.nuclear.lu.se/toi/nuclide.asp?iZA=840210
13. Tauber, C. et al: Heterogeneous nucleation onto monoatomic ions: Support for the Kelvin-Thomson theory. ChemPhysChem (2018). https://doi.org/10.1002/cphc.201800698
14. https://en.wikipedia.org/wiki/Ethanol.
15. Diemand, J., Angelil, R., Tanaka, K. K., Tanaka, H.: Large scale molecular dynamics simulations of homogeneous nucleation. J. Chem. Phys. **139**, 074309 (2013). https://doi.org/10.1063/1.4818639
16. Horsch, M., Miroshnichenko, S., Vrabec, J.: Steady-state molecular dynamics simulation of vapour to liquid nucleation with McDonald's daemon. J. Phys. Studies **13**, 4004 (2009). https://doi.org/10.30970/jps.13.4004
17. Maximoff, S. N., Salehi, A., Rostani, A. A.: Molecular dynamics simulations of homogeneous nucleation of liquid phase in highly supersaturated propylene glycol vapors. J. Aerosol Science **154**, 105743 (2021). https://doi.org/10.1016/j.jaerosci.2020.105743